\documentclass[a4paper]{jpconf}

\usepackage{graphicx}
\usepackage{amssymb}

\begin{document}

\title{The dust emission of high-redshift quasars}

\author{C Leipski$^1$, K Meisenheimer$^1$}

\address{$^1$ Max-Planck Institut f\"ur Astronomie (MPIA), K\"onigstuhl 17, D
-69117 Heidelberg, Germany}

\ead{leipski@mpia-hd.mpg.de}

\begin{abstract}

The detection of powerful near-infrared emission in high redshift
($z>5$) quasars demonstrates that very hot dust is present close to the
active nucleus also in the very early universe. A number of 
high-redshift objects even show significant excess emission in the rest 
frame NIR over more local AGN spectral energy distribution (SED) templates. In 
order to 
test if this is a result of the very high luminosities and redshifts, we 
construct mean SEDs from the latest SDSS quasar catalogue in combination 
with MIR data from the WISE preliminary data release for several redshift 
and luminosity bins. Comparing these mean 
SEDs with a large sample of $z>5$ quasars we could not identify any 
significant trends 
of the NIR spectral slope with luminosity or redshift in the regime 
$2.5<z \lesssim 6$ and 
$10^{45} < \nu {\rm L}_{\nu}(1350{\rm \AA}) \lesssim 10^{47}$\,erg/s. 
In addition to the NIR regime, our combined {\it Herschel} and {\it Spitzer} 
photometry provides full infrared SED coverage of the same sample of $z>5$ 
quasars. These observations reveal strong FIR emission 
(L$_{\rm FIR}$\,$\gtrsim$\,10$^{13}$\,L$_{\odot}$) in seven objects, 
possibly indicating star-formation rates of several thousand 
solar masses per year. The FIR excess emission has unusally high temperatures 
(T\,$\sim$\,65\,K) which is in contrast to the 
temperature typically expected from studies at lower redshift 
(T\,$\sim$\,45\,K). These objects are currently being investigated in 
more detail.

\end{abstract}

\section{Introduction}

The presence of dust seems to be a ubiquitous property of galaxies
throughout the observable universe. Even the most distant quasars at
$z\sim6$ show evidence for copious amounts of dust (e.g. Bertoldi et
al. 2003;  Wang et al. 2008; Leipski et al. 2010). 
This indicates the rapid metal enrichment of the interstellar medium within 
the first billion years after the big bang. Assuming that the observed 
far-infrared (FIR) emission of these
objects is powered by star-formation, the luminosies imply star formation 
rates of up to a few
thousand solar masses per year, possibly indicating the rapid
formation of early galactic bulges. 

For a more 
comprehensive picture of the dust emission at high redshifts
we are currently analyzing PACS (100+160\,$\mu$m) and SPIRE
(250+350+500\,$\mu$m) photometry of 71 quasars at $z>5$ that have been 
obtained as part of our
{\it Herschel} key project "The Dusty Young Universe".  Complementary
{\it Spitzer} data at shorter wavelengths were secured by our group to
enable the study of the full optical through infrared SED of a large sample 
of quasars in the early universe. 

In the course of this project we have 
now identified a number of high-$z$ ($z>5$) quasars with considerable FIR 
emission. They are clearly detected in our PACS and SPIRE observations 
and were previously unknown to have large infrared luminosities (Fig.\,1). 
In combination with our existing {\it Spitzer} photometry, the {\it Herschel} 
data allow us study the full SED of these objects in the rest frame
wavelength range $0.5-80\,\mu$m, which -- most importantly -- also
covers the FIR peak of the SED.

\begin{figure}
\centering
\includegraphics[scale=0.4]{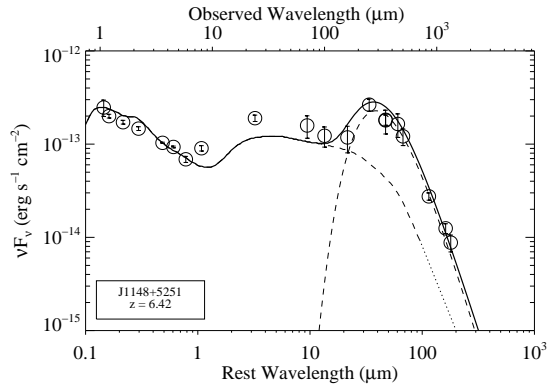}
\caption{SED for the $z=6.42$ quasar SDSS J1148+5251 including new
  {\it Herschel} data as well as information from the literature. {\it
    Left}: The quasar SED was fitted by a linear combination of the 
  SDSS quasar template (Richards et al. 2006) and a modified black body 
  (representing dust emission from star formation). Despite a reasonable 
  match in the rest frame UV/optical and in the FIR, there exist a clear 
  discrepancy between $\sim$\,$1-10\,\mu$m.
}
\end{figure}

\section{The signatures of hot dust in the rest frame near infrared}

Most powerful AGN are strong sources of infrared radiation. This
emission originates from dust heated by the accretion disk to various
temperatures depending on its distance from the nucleus and its
spatial distribution.
In such a configuration, the sublimation temperature of the dust 
($\sim$\,1500\,K) sets the inner 
boundary of the dust distribution in AGN. The dust at these
temperatures predominately re-emits the 
absorbed energy in the rest frame near infrared. Thus, the spectral energy 
distribution in this regime provides us with a tool to study the AGN dust 
only a 
few parsec out from the core.

Using {\it Herschel} data and archival 
information, we noticed that common AGN templates (like the SDSS
quasar template, Richards et al. 2006) do not match the amounts of 
hot dust emission between $\sim$\,1--10\,$\mu$m
 in the  UV/optical through sub-mm SED of the 
$z=6.42$ quasar SDSS\,J1148+5251 (Leipski et al. 2010; Fig.\,1). 
Interestingly, in the wavelength range where the
discrepancy is most apparent ($\sim$\,$1-5$\,$\mu$m), the SED is 
particularly sensitive to the
radial distribution of the innermost dust (however, this is not the only
parameter that can influence the shape of the SED in this region; 
see Schartmann et 
al. 2008, H\"onig \& Kishimoto 2010). 

The observed mismatch might be due to 
SDSS\,J1148+5251 being exceptional with a very high 
redshift ($z=6.42$ ) and having extreme luminosity 
(L$_{\rm bol}$\,$\sim$\,$10^{47}$\,erg/s). It is conceivable to 
assume that the dust distribution might be sensitive 
to the total luminosity of the heating
source (i.e. the accretion disk) or even the evolutionary state of the 
system (i.e. high-$z$ vs. low-$z$).

The SDSS template is largely 
dominated by quasars around a redshift of $z\sim2$. Only few objects 
at $z>3$ are included. Furthermore, while all objects have SDSS and 
{\it Spitzer}
IRAC data available, many objects lack NIR photometry (which was in
these cases determined from scaling the Elvis et al. (1994) quasar
template to the nearest observed data point). However, for a robust
comparison with our $z>5$ quasar SEDs -- in particular with an emphasis
on slope in the hot dust component -- we need a sample that provides
appropriate numbers of high-redshift quasars with observational data 
that cover the typical ``1\,$\mu$m inflection'' in the quasar SEDs and sample 
the onset of the hot dust emission. In addition, the indidviual SEDs
used to construct the SDSS template have been corrected for host
galaxy contributions (see Richards et al. (2006) for details) which
has the largest effect ($\sim$\,35\,\% host contribution) in the rest
frame NIR where the assumed elliptical host SED peaks.

In order to build template SEDs that cover higher redshifts at various
luninosities, we here use as a starting point the DR7 SDSS quasar
sample presented by Shen et al. (2011). These
authors also provide their quasar table matched with the
preliminary data release from the WISE (Wright et al. 2010) MIR survey. 
This results in 
$\sim$\,36000 quasars with SDSS optical phtometry and WISE MIR detections 
in three bands (3.4, 4.6, 12\,$\mu$m; we here do not consider the 
22\,$\mu$m band in which many objects are not detected). We further limited 
the objects to redshifts 
greater than 2.5 so that the WISE photometry samples the hot dust part of
the rest-frame quasar SED. From the remaining sources we then created mean
SEDs in a fairly straight forward manner by first de-redshifting the
sources into the rest frame, then interpolating them linearly in log
space onto a common wavelength grid, normalzing them at 1\,$\mu$m
(rest frame) and eventually calculating the mean for 
various subsamples. 
We then compared these mean SEDs with our sample
(Fig.\,2). Specifically, we searched for trends with increasing
redshift (at fixed luminosity) or with luminosity (at fixed redshift).
A few important findings are:
\begin{itemize}
\item We see quite a large spread in SED shapes for our sample as well as 
in the
  comparison samples after normalization, which results in substantial
  error bars (here: mean absolute deviation). Such a spread of quasar SEDs was
  also seen by e.g. Elvis et al. 1994 and Richards et al. 2006 in
  their studies.

\item No trend (within the errors) is observed when increasing the
  luminosity over 2 orders of magnitude ($\sim$\,10$^{45}$--10$^{47}$)
  while keeping the redshift bin fixed ($2.5<z<3$).

\item Equally no trend (within the errors) can be found with redshift 
  between $2.5<z<4$ when using only sources at comparable luminosity 
  to the high-$z$ objects.

\end{itemize}

This implies that the hot dust properties do not change significantly
with either redshift or luminosity for the parameter space considered
here. In fact, the diversity of individual quasar SEDs appears much
greater than any differences between the sample mean values.

We stress, however, that this study and the results presented here
have to be considered preliminary. We are currently looking into expanding
the observational data base for the quasars by including 
NIR photometery from the UKIDSS survey. This will allow us to expand this 
study to even lower redshifts while still appropriately sampling the hot-dust 
part of the SED. Also, no correction for host galaxy
contributions has been performed at this point which could be
especially important for the lower luminosity objects. Such a
correction could potentially influence the spectral shape in the NIR
where the stellar population of an elliptical galaxy of old or intermediate
age is brightest.

In fact, in a recent MIR interferometric study, Kishimoto et al. (2011) 
report a steepening of 
the radial dust distribution in the innermost few parsec with increasing 
luminosity. In the SEDs this 
would translate into a steeper NIR spectral slope for objects at higher
luminosity which indicates that a carefull accounting of the
host galaxy contribution may prove to be important for the less
luminous sources.

\begin{figure}
\centering
\includegraphics[scale=0.8]{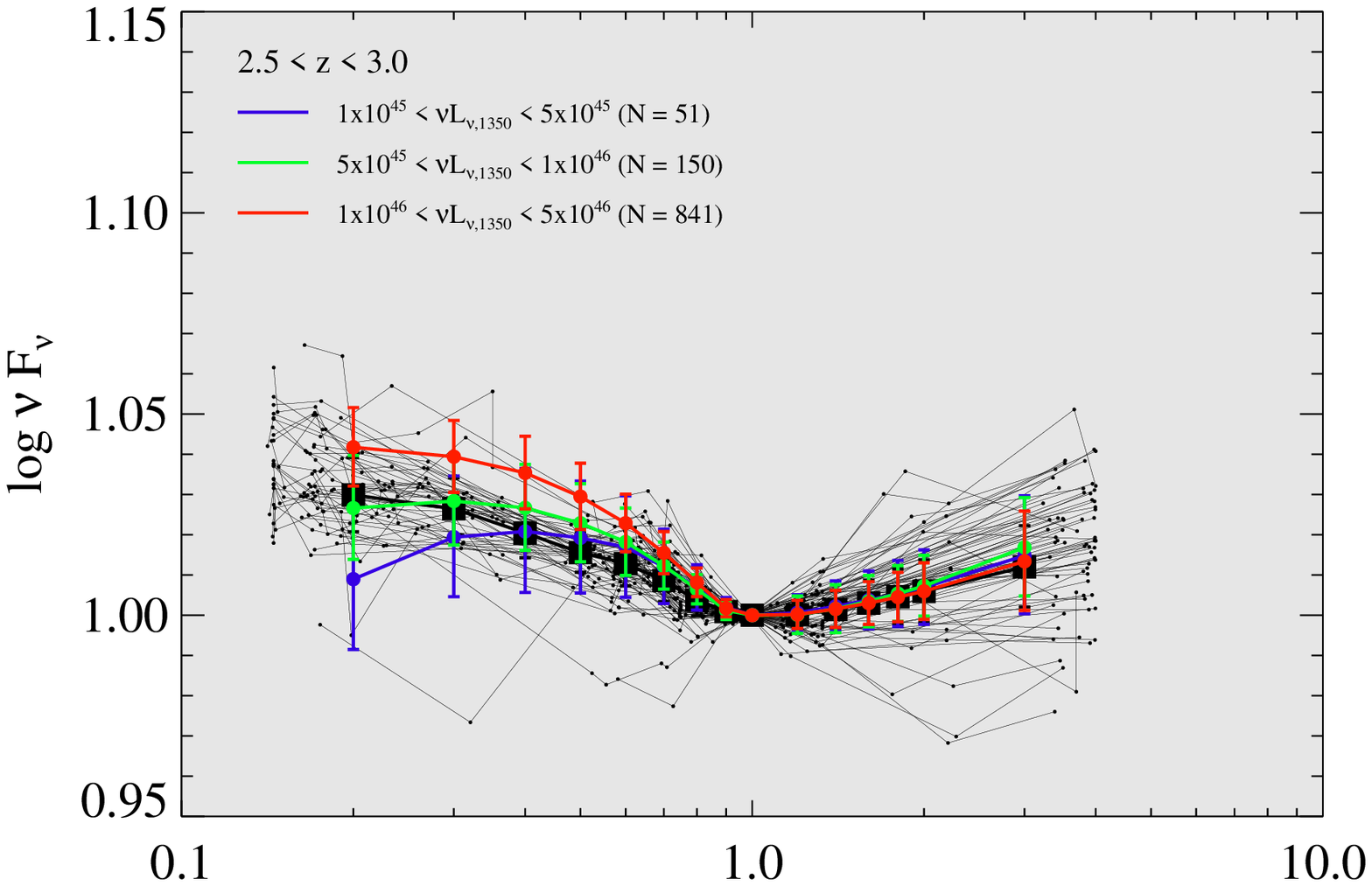}\\
\includegraphics[scale=0.8]{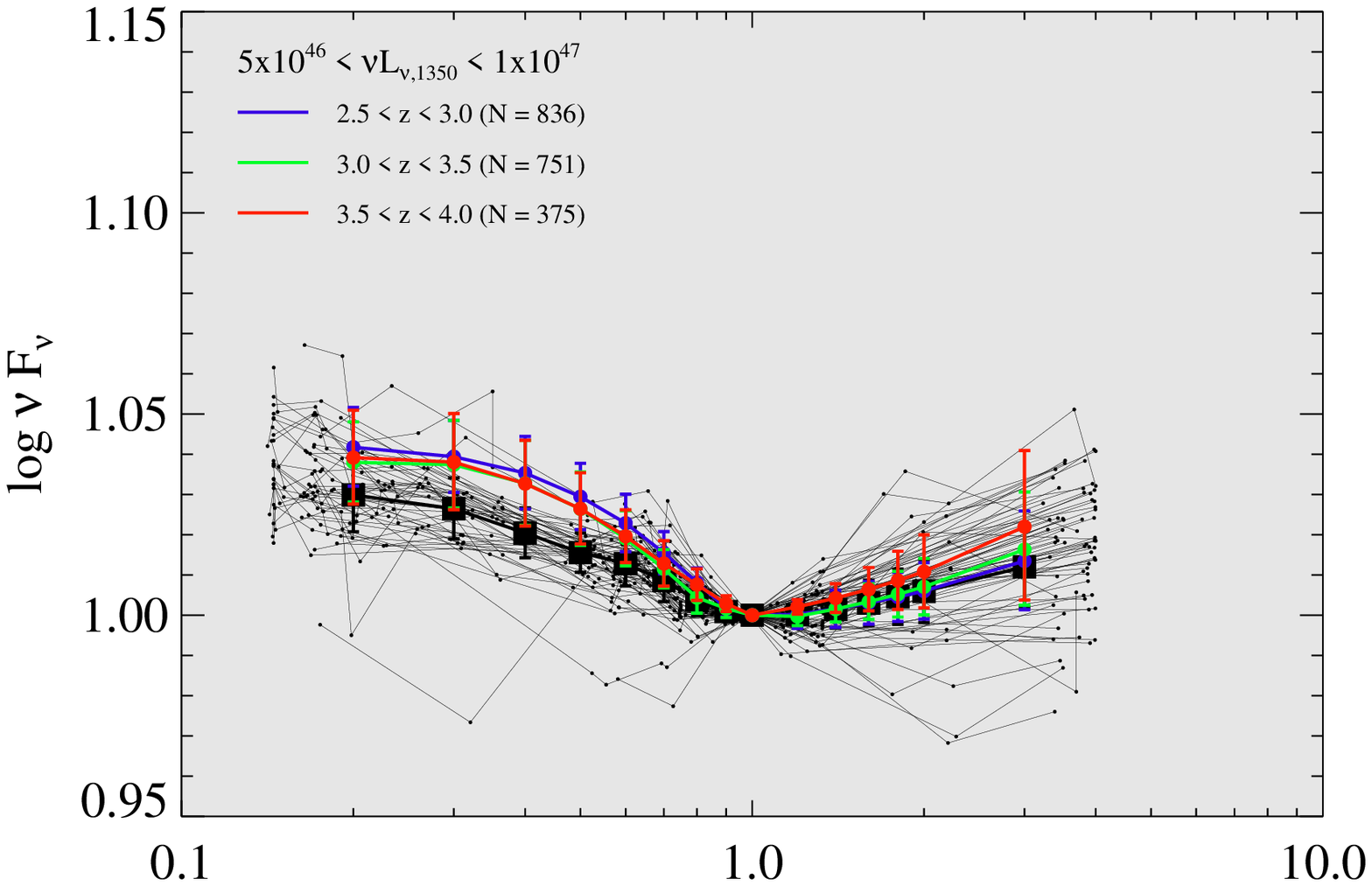}
\caption{Comparison of normalized SEDs. In both panels the black
  points indicate the observed photometry for our $z>5$ sample 
(here 68 individual sources are shown). The
  black squares represents a mean SED for our sample created in a
similar fashion as for the comparison SEDs. {\it Top}: Our sample
compared with mean SEDs for various luminosity intervals but a common
redshift regime $2.5<z<3.0$. We give the luminosity intervals and
number of objects used to create a specific SED in the plot. {\it
  Bottom}: Same as above, but now for a fixed luminosity, and creating
the mean SEDs for various redshift intervals.}
\end{figure}

\section{Far-infrared emission powered by star formation}

  About 30\% of the known luminous
$z\sim6$ quasars are detected  at 250\,GHz and/or in CO (e.g. Bertoldi
et al. 2003; Wang et al. 2008).  Such studies suggest that most of the
rest-frame far-infrared (FIR) emission comes from massive star
formation, possibly indicating the formation of early galactic
bulges. Thus, these objects signify an important stage in the
connection between the build-up of stellar mass and black hole growth.

However, for the vast majority of high-redshift objects we lack full 
FIR/sub-mm 
spectral energy distributions (SEDs). Far-infraed luminosities
(L$_{\rm FIR}$) and dust masses (M$_{\rm dust}$ ) are commonly
determined using single photometric measurements (typically at
250\,GHz) and applying standard values for the dust temperature 
as determined from lower redshift objects. But the questions remains how far 
these assumptions are appropriate for the high-redshift objects. 

From our new photometry providing full SED
information in the rest frame wavelength range $0.5-80\,\mu$m, we have
discovered seven quasars at $z>5$ with considerable FIR
excess emission, including SDSS\,J1148+5251 for which strong
  FIR and mm emission has been reported previously (e.g. Bertoldi et
  al. 2003, Beelen et al. 2006).  
In order to extract additional information from the
SEDs, we fitted the observed photometry with a linear combination of a
power-law in the UV/optical/NIR (emission from the accretion disk), a 
clumpy torus 
model (AGN heated dust from H\"onig \& Kishimoto (2010)), and 
a modified black body (starburst-powered excess dust emission).
The emmissivity value $\beta$ for this last component was fixed
to a value of 1.6 to be comparable to previous studies at high $z$. In
all cases the additional FIR component was required to describe the
observed SEDs at the longest wavelengths (see Fig.\,3 for a few examples). 
These SED fits also
allow us to determine quantative parameters: first, the scaling
of the torus dust emission can be converted into the bolometric
luminosity required to power the observed emission (H\"onig \&
Kishimoto 2010). For the sources considered here we find 
L$_{\rm bol}$\,=\,$1-5\times 10^{47}$\,erg/s using this
approach. These bolometric luminosities translate into dust
sublimation radii of $\sim$\,$3-5$\,pc. We also see that in the
combined fits, torus models with steep radial dust distributions are
preferred, indicating that most of the dust resides close to the
nucleus. Integrating only the FIR excess component between 8 and 
1000\,$\mu$m we determine luminosities of
$\gtrsim$\,10$^{13}$\,L$_{\odot}$. If we identify this component as being
powered by star formation then we observe star formation rates of the
order of a few times 10$^3$ solar masses per year in these early quasar 
host galaxies.
For SDSS\,J1148+5251 the star-formation rate we determine from the SED
fits is consistent with the results from a [CII] imaging study (Walter
et al. 2009). These spatially resolved observations also show that 
the bulk of the detected star formation takes place in the innermost 
1.5\,kpc of the host galaxy.

One interesting fact related to the FIR excess emission is
that the temperature of this component in all sources is
fairly high (T\,$\sim$\,65\,K), much higher than the 
T\,$\sim$\,45\,K typically found for lower redshift objects
(e.g. Beelen et al. 2006). This latter value is commonly used to 
extrapolate from the 250\,GHz measurements in high-redshift 
quasars (e.g. Wang et al. 2008). Since the determination of dust masses 
is quite sensitive to the temperature of the dust, the measured 
difference in FIR temperatures could potentially have an impact on the 
dust masses derived for the high-redshift quasars.

Our multi-wavelength imaging allows us to check for possible 
confusion with
other (lower redshift) FIR bright sources which could contribute in
the large SPIRE beams and mimick the higher FIR dust temperatures.
No such confusion problems could readily be identified, at least on
scales of a few arc seconds.
We might also have to consider quasar heating of the dust as a possible
soure to raise the temperature of FIR emitting dust. The very luminous
AGN puts out copious amounts of radiation which 
in principle might be able to heat dust in the host galaxy to the
temperatures we observe. In such a case, only some fraction of the FIR
luminosity will actually be powered by star formation and in our
calculations above we would overestimate the
star-formation rates in these objects. A more detailed analysis of these 
highly interesting sources is
currently underway (Leipski et al. 2012, in
preparation) and follow-up strategies have been initiated.

\begin{figure}
\centering
\includegraphics[scale=0.4]{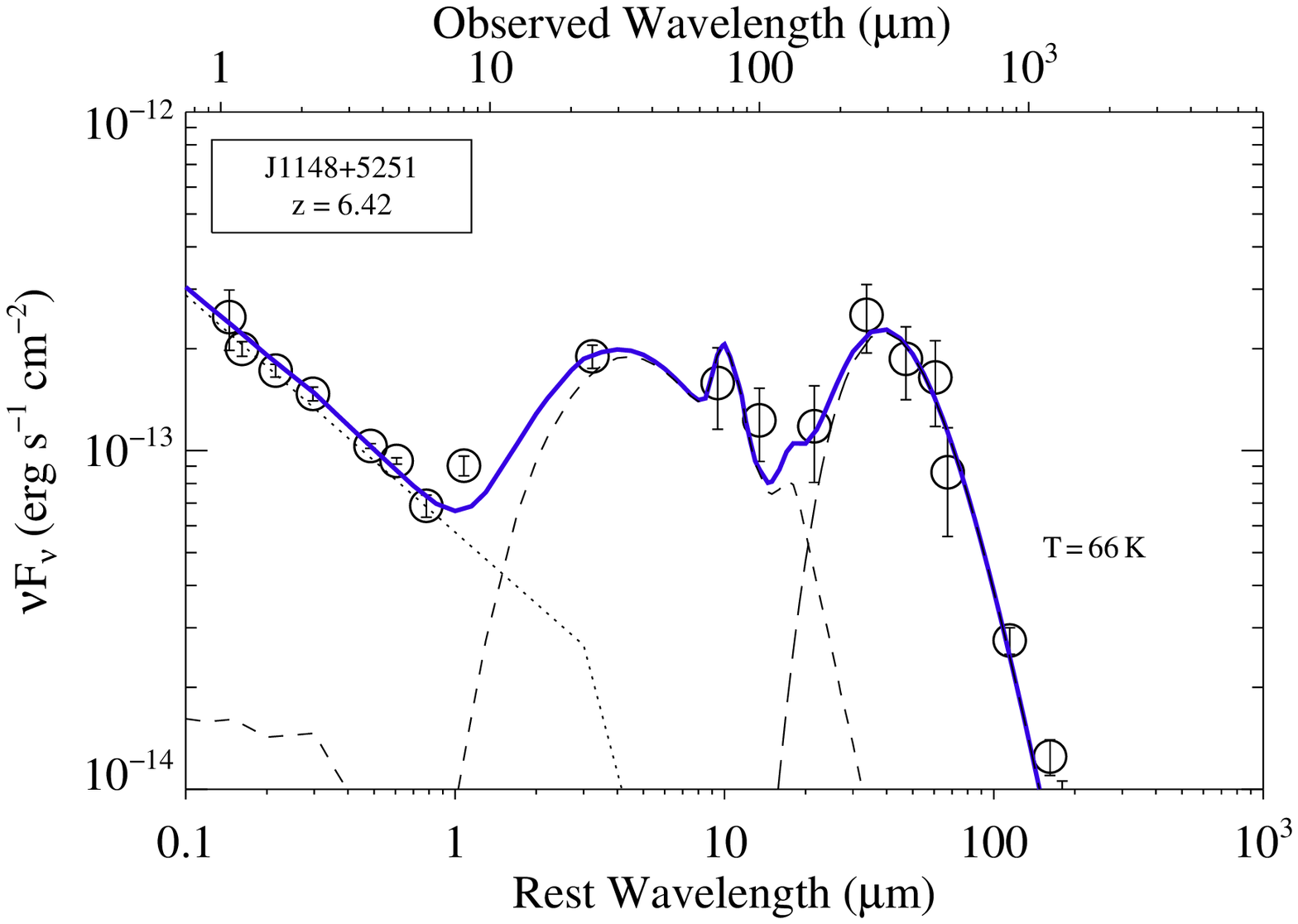}
\includegraphics[scale=0.4]{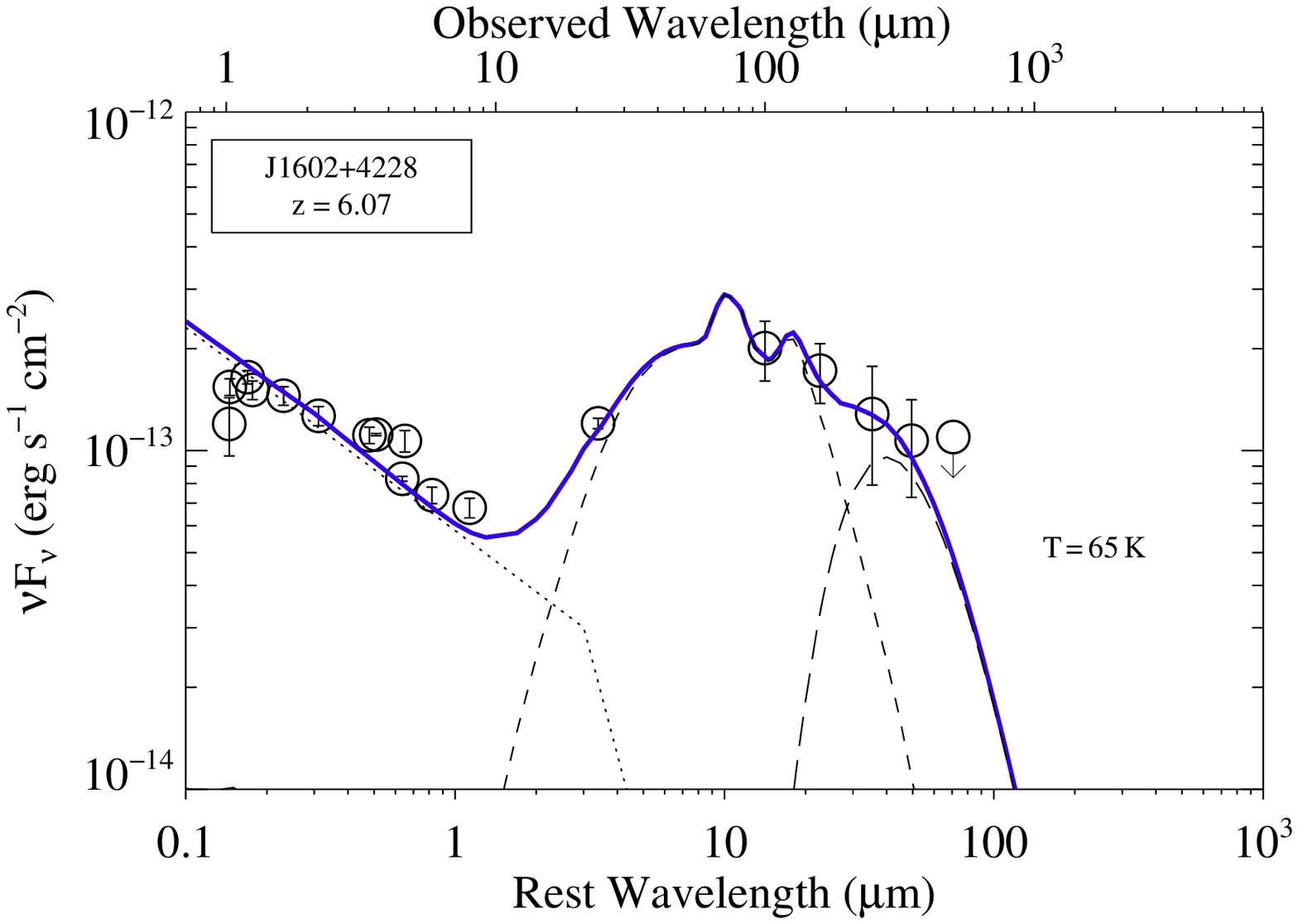}\\[1.0ex]
\includegraphics[scale=0.4]{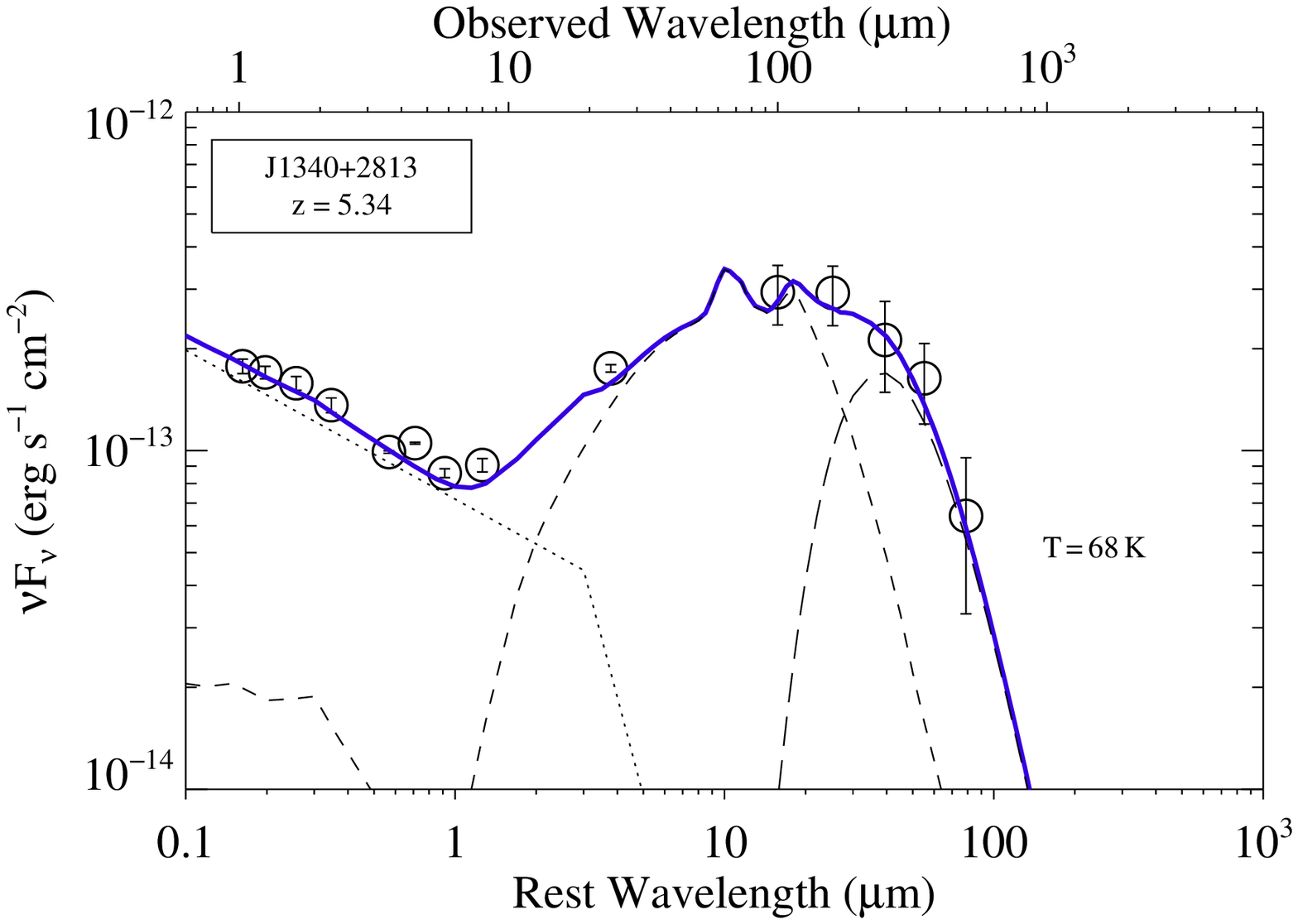}
\includegraphics[scale=0.4]{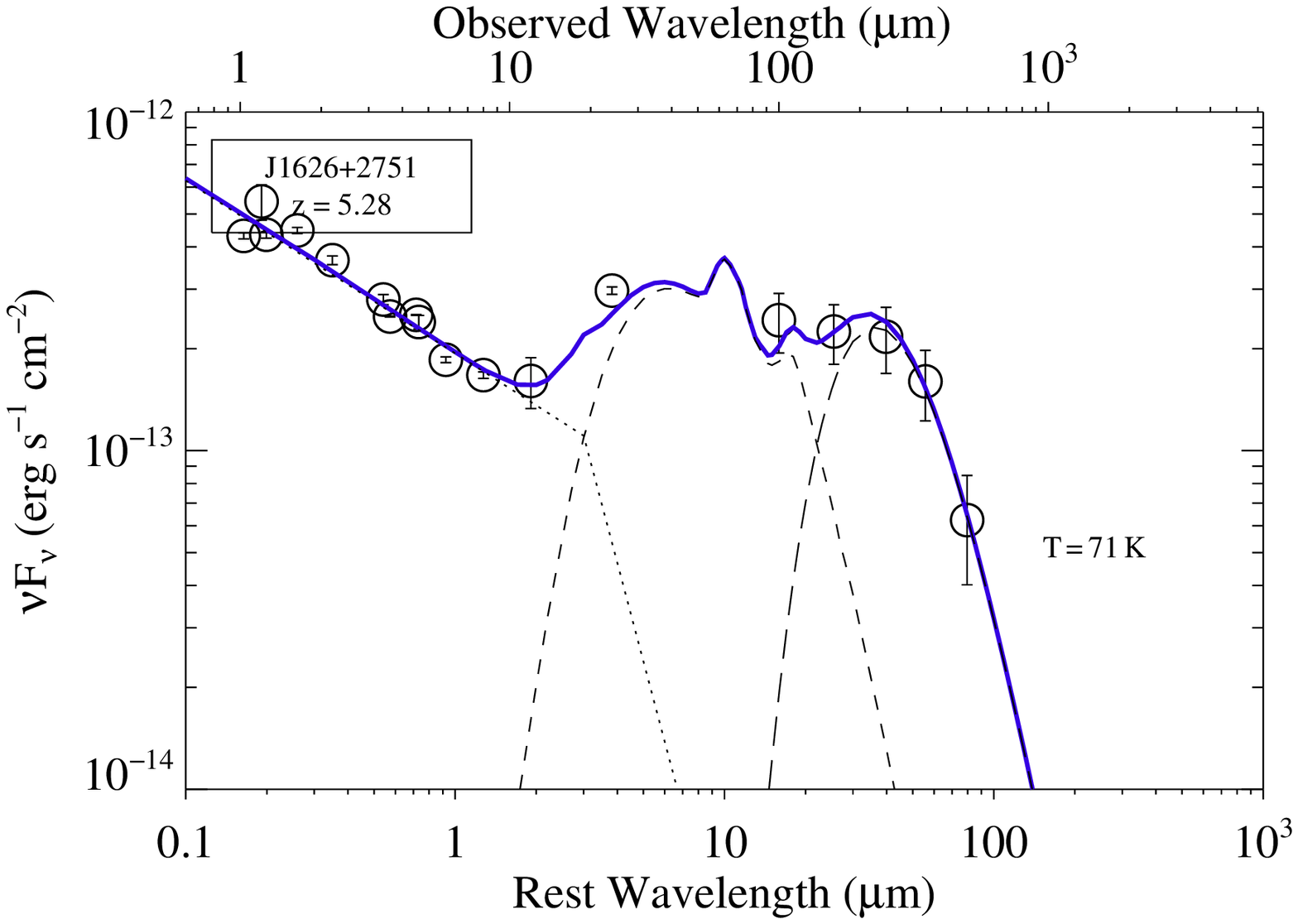}
\caption{Full optical through infrared SEDs of four sources
  showing FIR excess emission with high temperatures. In order to disentangle 
AGN and starburst powered emission, the photometry 
has been fitted by a linear combination of three components: a
power-law in the UV/optical/NIR (dotted line), a clumpy torus 
model (dashed line), and 
a modified black body (long-dashed line). The blue solid line shows
the total fit. For all sources shown here, the temperature of the FIR
component is T\,$\sim$\,65\,K.
Note also that compared to Fig.\,1 the AGN 
dominated MIR dust emission in  SDSS\,J1148+5251 is much better described 
by this model, as is the overlap 
region of the AGN and the starburst ``bump''.}

\end{figure}

\section{Summary}

We present first results from our {\it Herschel} and {\it Spitzer}
observations of $z>5$ quasars. We find that existing quasar templates
do not match the amount of hot dust emission observed in almost all
high-$z$ quasars. Using the latest SDSS quasar catalogue including
data from the WISE MIR survey we do not detect any significant trends
in the shape of the NIR SED when considering mean SEDs at various
redshifts and luminosities. Our full SED coverage in the infrared 
reveals considerable FIR excess emission in seven objects. 
If powered by star formation, the FIR luminosities determined from SED 
fitting indicate 
star-formation rates of thousands of solar masses per
year. Furthermore, the FIR emission in these objects shows unusually
high temperatures ($\sim$\,65\,K as opposed to 45\,K at lower $z$). 
The analysis on these aspects is ongoing.

\ack
We would like to thank M. Schartmann and S. H\"onig for providing 
their respective torus models as well as for their valuable insight 
into the modeling process.

\section*{References}

\begin{thereferences}

\item Beelen A, Cox P, Benford D~J et al.\ 2006,  {\it Astrophysical Journal}, {\bf 642}, 694 

\item Bertoldi F, Carilli C~L, Cox P et al.\ 2003,  {\it Astronomy \& Astrophysics}, {\bf 406}, L55 

\item Elvis M, Wilkes B~J, McDowell J~C et al.\ 1994, {\it Astrophysical Journal Supplement
  Series}, {\bf 95}, 1 

\item H\"onig S and Kishimoto M 2010,  {\it Astronomy \& Astrophysics}, {\bf 523}, 27 

\item Kishimoto M, H{\"o}nig S~F, Antonucci R et al.\ 2011, {\it Astronomy \& Astrophysics}, {\bf 536}, A78 

\item Leipski C, Meisenheimer K, Klaas U et al. 2010,  {\it Astronomy \& Astrophysics}, {\bf 518}, L34

\item Richards G~T, Lacy M, Storrie-Lombardi L~J et al.\ 2006,
  {\it Astrophysical Journal Supplement Series}, {\bf 166}, 470 

\item Walter F, Riechers D, Cox P et al. 2009, {\it Nature}, {\bf 457}, 699

\item Wang R, Carilli C~L, Wagg J et al.\ 2008, {\it Astrophysical Journal}, {\bf 687}, 848 

\item Wright E~L, Eisenhardt P~R~M, Mainzer A~K, et al.\
  2010, {\it Astronomical Journal}, {\bf 140}, 1868

\end{thereferences}

\end{document}